\documentclass[lettersize,journal]{IEEEtran}
\usepackage{amsmath,amsfonts}
\usepackage{algorithmic}
\usepackage{algorithm}
\usepackage{array}
\usepackage[caption=false,font=normalsize,labelfont=sf,textfont=sf]{subfig}
\usepackage{textcomp}
\usepackage{stfloats}
\usepackage{url}
\usepackage{verbatim}
\usepackage{graphicx}
\usepackage{cite}
\usepackage{multicol, multirow}
\usepackage{color}
\definecolor{navy}{rgb}{0, 0, 0.52}
\definecolor{royal}{rgb}{0.25, 0.42, 0.88}
\definecolor{medium_blue}{rgb}{0, 0, 0.8}
\usepackage{xpatch}
\makeatletter
\ExplSyntaxOn
\cs_new:Npn \bibColoredItems #1#2
  {
    \clist_map_inline:nn {#2} { \cs_new:cpn {bib@colored@##1} {#1} } 
  }
\ExplSyntaxOff

\newcommand\bib@setcolor[1]{%
  \ifcsname bib@colored@#1\endcsname
    \expandafter\color\expandafter{\csname bib@colored@#1\endcsname}
  \else
    \normalcolor
  \fi
}

\xpatchcmd\@bibitem
  {\item}
  {\bib@setcolor{#1}\item}
  {}{\PatchFailed}

\xpatchcmd\@lbibitem
  {\item}
  {\bib@setcolor{#2}\item}
  {}{\PatchFailed}
\makeatother

\begin{document}

\title{Optical Integrated Sensing and Communication: Architectures, Potentials and Challenges}

\author{Yunfeng~Wen,
Fang~Yang,~\IEEEmembership{Senior~Member,~IEEE},
Jian~Song,~\IEEEmembership{Fellow,~IEEE},
and~Zhu~Han,~\IEEEmembership{Fellow,~IEEE}
\thanks{This work was supported by National Key Research and Development Program of China under Grant 2022YFE0101700. \emph{(Corresponding author: Fang~Yang.)}}
\thanks{Yunfeng~Wen and Fang~Yang are with the Department of Electronic Engineering, Tsinghua University, Beijing 100084, P. R. China, and also with the Key Laboratory of Digital TV System of Shenzhen City, Research Institute of Tsinghua University in Shenzhen, Shenzhen 518057, P. R. China (e-mail: wenyf22@mails.tsinghua.edu.cn; fangyang@tsinghua.edu.cn).}
\thanks{Jian Song is with the Department of Electronic Engineering, Tsinghua University, Beijing 100084, P. R. China, and also with the Shenzhen International Graduate School, Tsinghua University, Shenzhen 518055, P. R. China (e-mail: jsong@tsinghua.edu.cn).}
\thanks{Zhu Han is with the Department of Electrical and Computer Engineering, University of Houston, Houston, TX 77004 USA, and also with the Department of Computer Science and Engineering, Kyung Hee University, Seoul 446-701, South Korea (e-mail: hanzhu22@gmail.com).}
}

\maketitle

\begin{abstract}
Integrated sensing and communication (ISAC) is viewed as a crucial component of future mobile networks and has gained much interest in both academia and industry. Similar to the emergence of radio-frequency (RF) ISAC, the integration of free space optical communication and optical sensing yields optical ISAC (O-ISAC), which is regarded as a powerful complement to its RF counterpart. In this article, we first introduce the generalized system structure of O-ISAC, and then elaborate on three advantages of O-ISAC, i.e., increasing communication rate, enhancing sensing precision, and reducing interference. Next, waveform design and resource allocation of O-ISAC are discussed based on pulsed waveform, constant-modulus waveform, and multi-carrier waveform. Furthermore, we put forward future trends and challenges of O-ISAC, which are expected to provide some valuable directions for future research.
\end{abstract}

\section{Introduction}
Future mobile networks are anticipated to provide both large communication capacities and high-precision sensing abilities simultaneously, which is regarded as a key enabler for numerous applications, e.g. intelligent transportation systems (ITS), human-computer interactions~\cite{RinchiLiDARHAR2023}, and the Internet of Things (IoT). However, the isolated development of sensing and communication has resulted in inefficient spectrum and hardware resource utilization, which have been increasingly scarce in recent years. Consequently, the concept of integrated sensing and communication (ISAC) is motivated in the evolution of future mobile networks such as the sixth-generation (6G) systems.

The scope of ISAC encompasses the coexistence, cooperation, and co-design of communication and sensing. The independent operation of communication and sensing is maintained by coexistence, which reduces their mutual interference. Furthermore, performance gains can be obtained through cooperation, where communication and sensing share valuable information with each other. Finally, the system architecture and parameters can be jointly designed for communication and sensing, which achieves the global optimum for the whole system. While coexistence is cost-effective, cooperation and co-design are indispensable for superior performances in both communication and sensing.

Up to now, numerous research has been conducted on the radio-frequency (RF) ISAC, spanning several areas like waveform design, networking, resource allocation, and information theory. Among these areas, waveform design has gained much interest as a fundamental aspect of ISAC. The ISAC waveform can be designed in a communication-centric or sensing-centric approach and can also be conceived from the ground up~\cite{LiuISAC2022}. Towards this end, an enormous amount of RF-ISAC waveforms have been designed based on index modulation, linear frequency modulation (LFM), and orthogonal frequency division multiplexing (OFDM).

In the context of ISAC, the optical band exhibits striking similarities to the RF band. First, well-established RF-ISAC waveforms like LFM and OFDM can be transformed into optical waveforms, enabling optical devices like laser radars to provide communication and active sensing abilities simultaneously. Second, both free-space optical (FSO) communication and optical sensing can adopt the intensity modulation and direct detection (IM/DD) scheme, allowing for the sharing of optical front-ends and baseband units. Third, the principles of information sharing and joint optimization apply to both RF and optical waveform, through which cooperation between communication and sensing can be achieved. However, significant differences still exist between FSO systems and their RF counterparts. Specifically, the unique characteristics of optical front-ends and channels offer promising advantages for communication and sensing, thereby serving as a powerful complement to RF-ISAC. In consequence, the integration of FSO communication and optical sensing gives rise to the concept of optical ISAC (O-ISAC).

This article investigates the basic architectures, opportunities, and challenges of O-ISAC to explore its vast potential. First, we outline the generalized system structure and then highlight three competitive advantages of O-ISAC. Second, waveform design and resource allocation aspects of O-ISAC are elaborated comprehensively based on widely adopted schemes such as pulsed waveform, constant-modulus waveform, and multi-carrier waveform. Third, we discuss the future trends of O-ISAC in conjunction with other emerging technologies, while also introducing the challenges to be addressed. We expect this article to provide valuable directions for future research in the field of O-ISAC.

\begin{figure*}[tp]
    \begin{center}
        \includegraphics[width=0.96\textwidth]{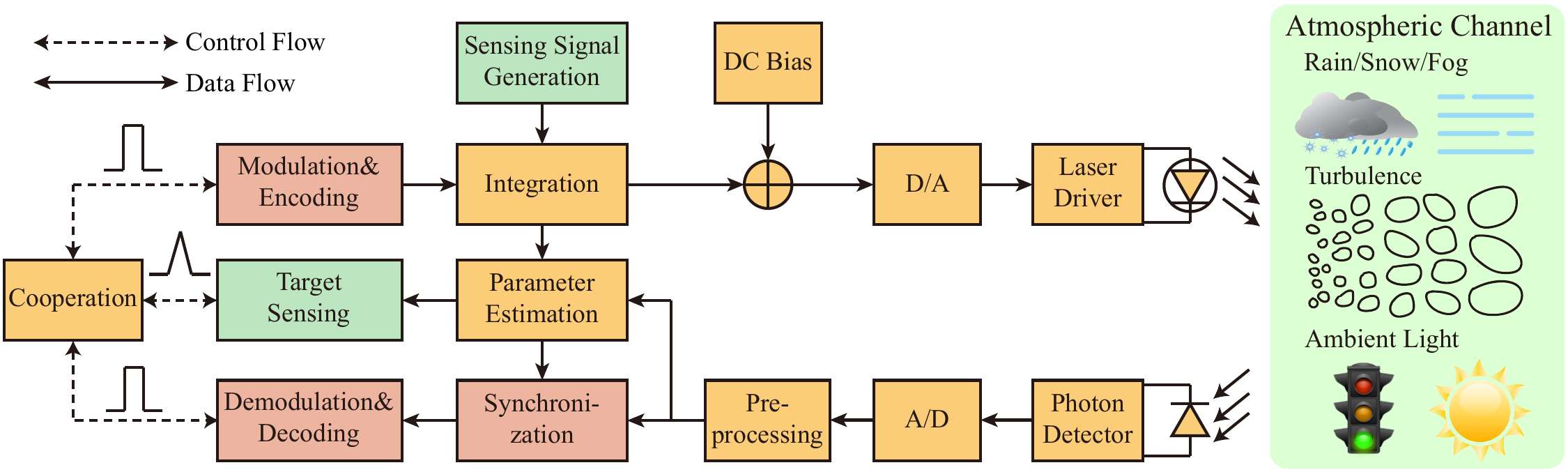}
        \caption[short]{Generalized system structure of O-ISAC. The blocks in yellow, red, and green are those shared by communication and sensing, specific to communication, and specific to sensing, respectively.}
        \label{SystemStructure}
    \end{center}
\end{figure*}

\begin{table*}[bp]
    \centering
    \renewcommand\arraystretch{1.2}
    \caption{Comparisons between RF-ISAC and O-ISAC}
    \begin{tabular}{c|c|c|c|c}
        \hline \hline
        \multicolumn{2}{c|}{\multirow{2}{*}{Characteristics}} & \multicolumn{2}{c|}{RF-ISAC} & \multirow{2}{*}{O-ISAC} \\ \cline{3-4}
        \multicolumn{2}{c|}{} & WiFi & Mm-wave & \\ \hline
        \multirow{5}{*}{Physical properties} & Frequency & 2.4 GHz, 5 GHz & 30 GHz - 300 GHz & 28.3 THz - 845 THz \\ \cline{2-5}
        & Amplitude & \multicolumn{2}{c|}{Complex} & Real and non-negative \\ \cline{2-5}
        & (De)modulation & \multicolumn{2}{c|}{Coherent} & IM/DD \\ \cline{2-5}
        & Channel & LoS \& NLoS & Mostly LoS & LoS \\ \cline{2-5}
        & Interference & Severe & Moderate & Slight \\ \hline
        \multirow{2}{*}{Communication metrics} & Communication distance & $<$100 m & $<$100 m & $<$1 km \\ \cline{2-5}
        & Achievable data rate & $\sim$100 Mbps & $\sim$Gbps & $\sim$Gbps \\ \hline
        \multirow{3}{*}{Sensing metrics} & Sensing distance & $<$100 m & $<$1 km & $<$1 km \\ \cline{2-5}
        &Distance resolution & $\sim$0.1 m & $\sim$cm & $\sim$cm \\ \cline{2-5}
        &Angle resolution & N/A & $\sim$1 mrad & $\sim$1 {\textmu}rad \\ \hline\hline
    \end{tabular}
    \label{Comparison}
\end{table*}

\section{System Structure and Advantages of O-ISAC}
\begin{figure*}[tp]
    \begin{center}
        \includegraphics[width=0.96\textwidth]{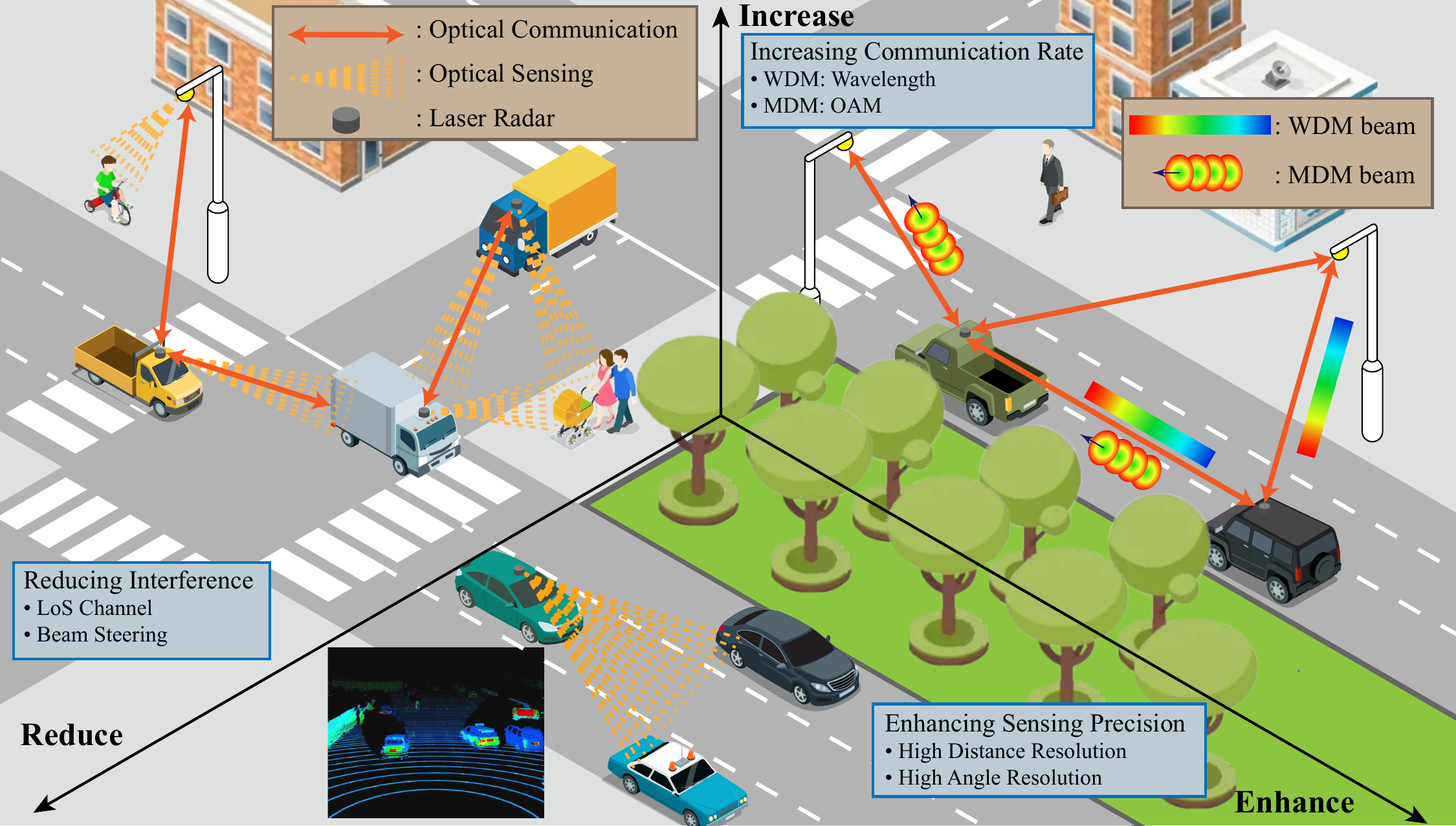}
        \caption[short]{Three advantages of O-ISAC: increasing communication rate, enhancing sensing precision, and reducing interference.}
        \label{Advantages}
    \end{center}
\end{figure*}

O-ISAC integrates FSO communication and optical sensing in an individual system, where the appealing characteristics of optical band bring three advantages, i.e., increasing communication rate, enhancing sensing precision, and reducing multi-user interference (MUI), as illustrated in Fig.~\ref{Advantages}. In this section, the generalized system structure of O-ISAC is first outlined in Section~\ref{SystemStructureSec}. Subsequently, the three advantages of O-ISAC are further elaborated in Sections~\ref{AdvantagesComm}, \ref{AdvantagesSensing}, and \ref{AdvantagesInterference}, respectively.

\subsection{System Structure of O-ISAC}\label{SystemStructureSec}
The generalized system structure of O-ISAC is illustrated in Fig.~\ref{SystemStructure}. On the transmitter side, the communication data is first encoded and modulated, and then integrated with the sensing signal to generate the O-ISAC signal. Then, the O-ISAC signal is direct-current (DC)-biased and transmitted to free space by the laser diode. 

The O-ISAC signal propagates in an atmospheric channel and is impaired by attenuation, turbulence, and ambient light. While atmospheric attenuation and turbulence are usually considered at the receiver, the ambient light can be suppressed by incorporating a closed-loop circuit or photon coincidence detection~\cite{MaikSPADLiDAR2018}. Additionally, a portion of the O-ISAC signal is reflected by the target, whose spatial information is then embedded into the reflected signal and then extracted by the sensing receiver.

On the receiver side, photon detector (PD) and analog-to-digital converter (A/D) are common modules shared by both communication and sensing receivers, followed by pre-processing units that suppress the ambient light and amplify the received signal. After the estimation of necessary parameters like the time of flight (ToF), target information like the distance and velocity can be attained by the sensing receiver, and the communication receiver can conduct synchronization, demodulation, and decoding.

Enlightened by the idea of perceptive mobile networks~\cite{ZhangPMN2020}, the implementation of O-ISAC becomes cost-effective on existing optical communication or sensing devices. Note that if the communication blocks in Fig.~\ref{SystemStructure} are omitted, the working principle of the O-ISAC system is the same as a laser radar, i.e., the optical counterpart of a radar. Meanwhile, by modulating the transmitted optical signal, the laser radar can also conduct FSO communication with other existing FSO receivers. Therefore, implementing O-ISAC on a laser radar or other optical sensors is more appealing than developing a new system from the ground up, just like the deployment of RF-ISAC based on existing base stations.

However, even if O-ISAC resembles RF-ISAC in the ideologies of hardware reuse and unified waveform design, the main differences between them consist in optical front-ends and channels. For instance, RF-ISAC signal propagates through a dispersive channel where frequency-selective fading is non-negligible. On the contrary, FSO channels are usually recognized as atmospheric-turbulence channels, where atmospheric attenuation and turbulence are the main factors affecting the O-ISAC signal. Moreover, large antenna arrays in RF-ISAC are also substituted by laser diodes and PDs to transceive O-ISAC signal. To further understand these distinctions, comparisons between O-ISAC and well-established RF-ISAC schemes, i.e. WiFi and mm-wave, are described in Table~\ref{Comparison}. While the information on RF-ISAC can be found in~\cite{LiuISAC2022}, we elaborate on O-ISAC in the following subsections.

\subsection{Increasing Communication Rate}\label{AdvantagesComm}
Contrary to the crowded RF spectrum, the optical band offers a broad unlicensed spectrum to conduct high-rate communication, and the communication rate of an experimental prototype utilizing 1550-nm laser can reach 100~Gbps at a distance of about 700~m~\cite{WalshDemonstrationFSOC2022}. Additionally, benefiting from the excellent monochromaticity, collimation, and coherence of laser beams, multiple independent data channels can be transmitted in parallel with multiplexing schemes like wavelength division multiplexing (WDM) and mode division multiplexing (MDM), through which the communication rate can be further enhanced.

\subsubsection{Wavelength Division Multiplexing}
The linewidth of laser diodes can be less than 1~nm, while their wavelength span can be in the order of 50~nm. Therefore, different optical signal can be divided finely in the wavelength domain, which lays the foundation for WDM. Considering the FSO data transmission system in~\cite{DuttaFSOOAM2022}, 16~modes with a spacing of 40~GHz (192.78~THz-193.38~THz) are combined at the transmitter, and the receiver harnesses tunable optical bandpass filters to separate different signal. In consequence, 16 channels of communication data can be transmitted simultaneously, which increases the communication rate by 16 times.

\subsubsection{Mode Division Multiplexing}
Multiple data channels can also be loaded on different spatial modes, among which orbital angular momentum (OAM) is a common set of orthogonal modes. Laser beams with helical phase front can be quantified as different OAM states, and are orthogonal while propagating coaxially. Moreover, OAM-based MDM can be adopted concurrently with WDM to further increase the communication rate, and therefore the FSO data transmission system in~\cite{DuttaFSOOAM2022} utilizes 16-WDM and 4-OAM beams simultaneously to provide 64~parallel channels in total.

\subsection{Enhancing Sensing Precision}\label{AdvantagesSensing}
As widely utilized optical sensors, laser radars are capable of detection, ranging, imaging, and recognition. Compared to their RF counterparts, laser radars can achieve a higher resolution and obtain more delicate spatial information under the same working principle, thanks to the wide bandwidth and small divergence angle of laser beams.

\subsubsection{High Distance Resolution}
The distance resolution is inversely proportional to the signal bandwidth. Since the optical band provides a wider bandwidth than that of the RF band, O-ISAC is anticipated to achieve a higher distance resolution than RF-ISAC. For instance, a frequency-modulated continuous-wave laser radar with downlink communication capability is demonstrated in~\cite{XuFMCWLidar2020}, where an LFM signal with a bandwidth of 5~GHz is adopted to achieve a distance accuracy of $\pm$2.2~cm under a signal-to-noise ratio (SNR) of 22.5 dB. Although a higher resolution is achieved by O-ISAC, it is not the only bottleneck in sensing, since the bandwidth of A/D, sampling rate, and noise also affect the sensing precision. Towards this end, signal processing techniques like subspace methods can be adopted to enhance the sensing performance within limited hardware and bandwidth resources.

\subsubsection{High Angle Resolution}
The divergence angle of an electro-magnetic wave declines as its wavelength decreases under the same aperture. For instance, an 80-{\textmu}rad divergence angle can be achieved by the 1550-nm laser ranging system in~\cite{DuPhotonCountingLaserRanging2018}, which is much smaller than the milliradian-level divergence angle of mm-wave ISAC. Thereby, laser radars with collimated beams are capable of distinguishing targets that are probably located in the same angle unit of conventional radars, thus achieving a higher angle resolution than their RF counterparts. However, a higher angle resolution also demands a more delicate acquisition, tracking, and pointing mechanism, which poses more stringent requirements for both hardware and algorithms.

\begin{figure*}[tp]
    \begin{center}
        \includegraphics[width=0.92\textwidth]{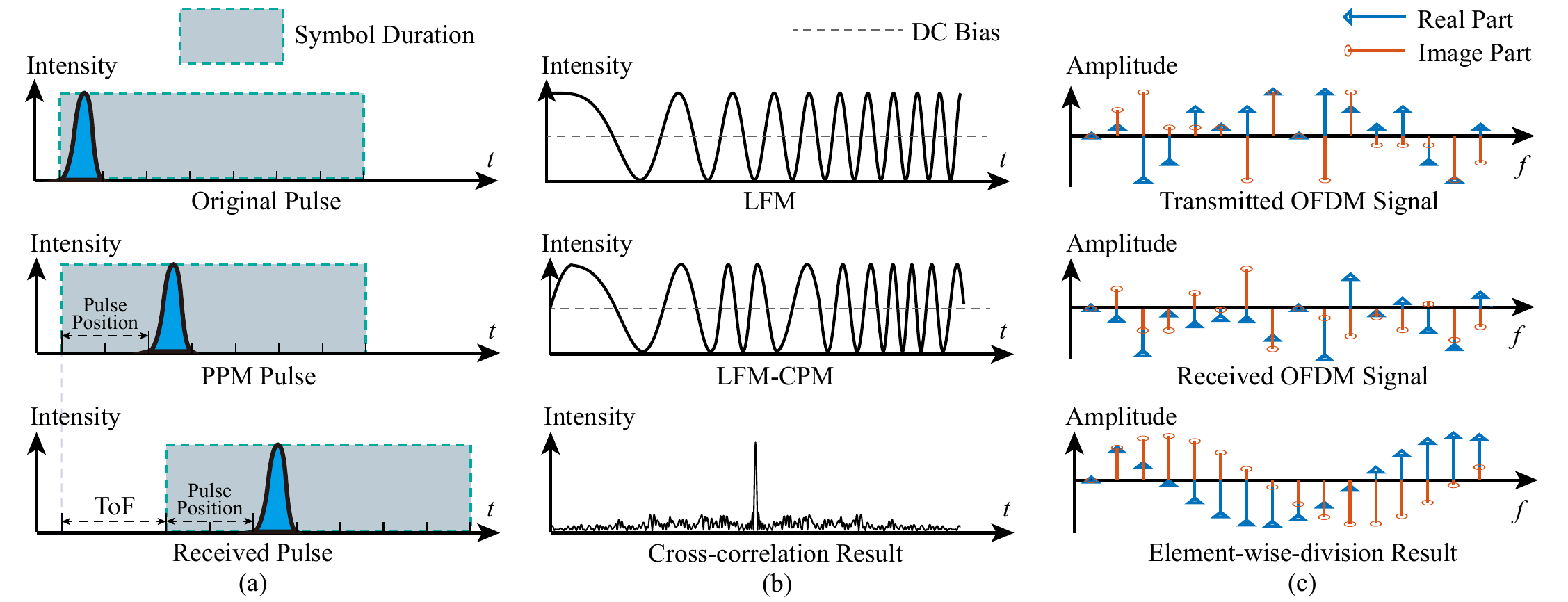}        
        \caption[short]{Waveform design for O-ISAC. (a) Pulsed waveform. (b) Constant-modulus waveform. (c) Multi-carrier waveform in the frequency domain.}
        \label{WaveformDesign}
    \end{center}
\end{figure*}

\subsection{Reducing Interference}\label{AdvantagesInterference}
One of the urgent tasks for ISAC is to mitigate the interference between communication and sensing. The optical band provides even narrower beams than the millimeter-wave with the same aperture, and therefore different user devices and access points can be separated finely in the spatial domain. Furthermore, O-ISAC generally depends on LoS channels, which eliminates the potential interference as optical beams cannot penetrate opaque obstacles. Meanwhile, beam steering and beam management techniques can also be introduced to reduce interference.

\subsubsection{LoS Channel}
Regardless of the negligible penetration for opaque obstacles, the path loss of an optical channel is primarily determined by the propagation distance and the reflectance factor. The reflectance factor varies dramatically with the material and roughness of the reflector, ranging from 0.2 for rough concrete to 0.8 for silvered mirrors. However, even for a high reflectance factor, the multiplicative attenuation brought by the extra propagation distance still limits the power of none-LoS (NLoS) channels. Therefore, the interference that depends on NLoS channels can be ignored in general, which yields superior anti-interference abilities for O-ISAC.

\subsubsection{Beam Steering}
Similar to the beamforming techniques in RF-ISAC, O-ISAC can also exploit beam-steering techniques to achieve spatial division multiplexing. By dynamically adjusting the direction and shape of laser beams, O-ISAC systems can avoid interference from unexpected sources and focus the beams precisely on the targets. While most existing beam-steering systems are bulky, fragile, and expensive, an acousto-optic beam-steering technique is reported in~\cite{LiAcoustoOpticLiDAR2023}, which achieves 2-mrad angular resolution in a chip-scale system. Towards this end, it is worth studying to integrate these miniaturized beam-steering system into O-ISAC for superior communication and sensing performances.

\section{Waveform Design for O-ISAC}
Waveform design is recognized as the foundation of ISAC, which focuses on designing an individual waveform that provides communication and sensing abilities simultaneously. Extensive research has been conducted on the waveform design for RF-ISAC, ranging from non-overlapped resource allocation to fully joint waveform design. However, an O-ISAC waveform based on the IM/DD scheme is restricted to being real and non-negative, and therefore most of the existing RF-ISAC waveforms are not readily applicable to O-ISAC, as they are complex in nature. To address this issue, several modifications have been proposed to make RF-ISAC waveform compatible with O-ISAC, like pulsed waveform, constant-modulus waveform, and multi-carrier waveform. Subsequently, waveform optimization and resource allocation can be further conducted for the prototype waveform to achieve the global optimal waveform design.

\begin{figure*}[tp]
    \begin{center}
        \includegraphics[width=0.96\textwidth]{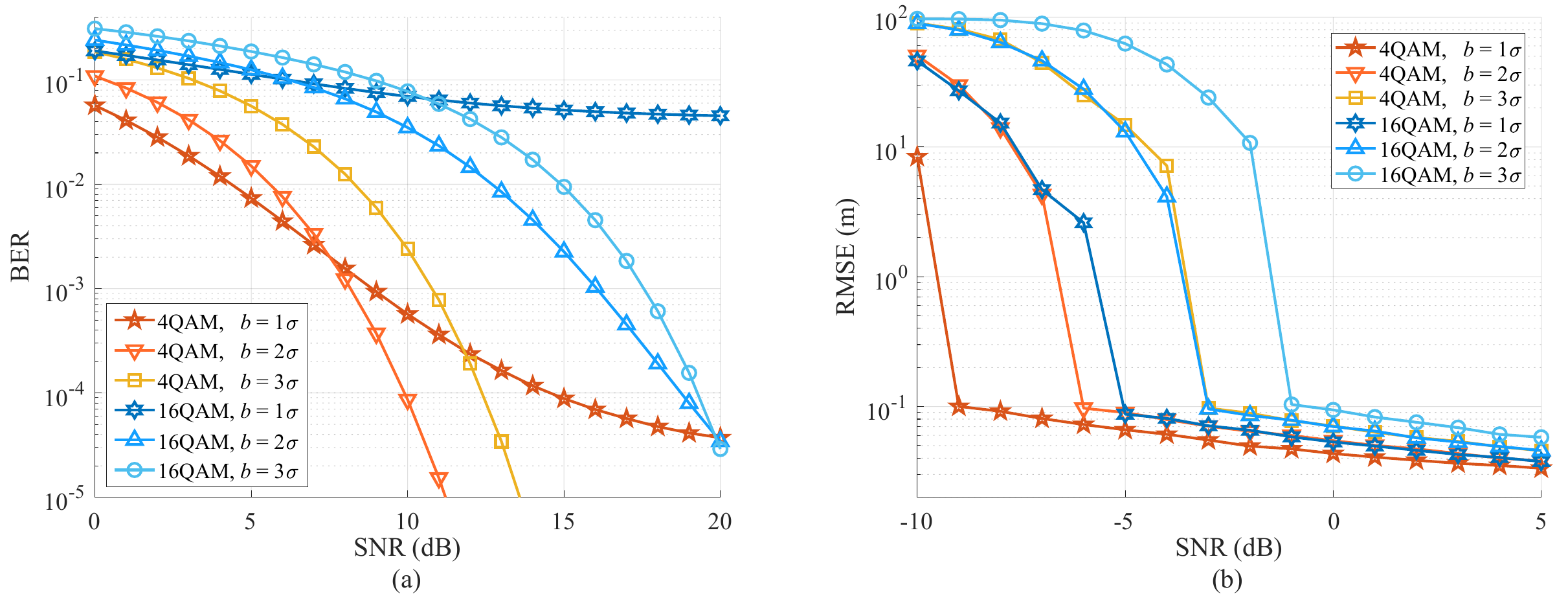}        
        \caption[short]{Numerical results of O-ISAC based on DCO-OFDM. The SNR is defined as the ratio between the transmitted power and the noise power at the receiver, while the attenuation of the channel is normalized. (a) BER versus SNR. (b) RMSE for target distance versus SNR.}
        \label{Simulation}
    \end{center}
\end{figure*}

\subsection{Pulsed Waveform}
The pulsed waveform is broadly embraced by both optical wireless communication and optical sensing owing to its high energy efficiency and cost-effective implementation. Pulse position modulation (PPM), which encodes communication data onto pulses in specific binary slots, is proven to be optimal in terms of energy efficiency. Moreover, distance information can also be extracted from the ToF of pulses once they are reflected by targets, which is a common working principle for pulsed laser radars. 

Since the position of pulses holds the information of communication data and target distance simultaneously, an O-ISAC scheme based on PPM is appealing, as illustrated in Fig.~\ref{WaveformDesign}(a). For instance, a pulse sequence sensing and pulse position modulation (PSS-PPM) scheme is proposed in~\cite{WenPSSPPM2023}, which modulates the position of pulses spread by m-sequences. Numerical simulations indicate that PSS-PPM can provide simultaneous communication and sensing abilities, even in the presence of severe MUI.

\subsection{Constant-modulus Waveform}\label{wav:constant_modulus}
The constant-modulus waveform is a subset of the subcarrier intensity modulation (SIM), where data is first modulated onto an electronic signal and then utilized to modulate the intensity of an optical signal~\cite{LiuPPMMSKSIM2015}. To obtain a real and non-negative optical signal, DC bias is necessary to guarantee that the negative part is not clipped. However, the constant-modulus characteristics reduce the demand for DC bias, and constant-modulus waveform remains energy-efficient compared with non-constant-modulus waveforms like pulse amplitude modulation or quadrature amplitude modulation (QAM).

One example for the constant-modulus waveform is the combination of continuous phase modulation (CPM) and LFM, as displayed in Fig.~\ref{WaveformDesign}(b). On the transmitter side, the complex baseband LFM-CPM signal is modulated to an intermediate frequency, whose real part is then DC-biased and utilized to modulate the intensity of the laser. On the receiver side, the Hilbert transform and down-conversion are adopted to recover the baseband signal. Afterwards, the sensing receiver calculates the cross-correlation to obtain a maximum-likelihood estimation of the target distance. Meanwhile, the communication receiver multiplies the baseband signal with the conjugate LFM signal for de-chirp, followed by the Viterbi decoding. When the sensing precision is set as a constraint, the LFM-CPM waveform is demonstrated to achieve a higher spectral efficiency than conventional CPM waveforms~\cite{WenFSOLFMCPM2023}.

\subsection{Multi-carrier Waveform}\label{wav:multi_carrier}
The multi-carrier waveform, as another subset of SIM, sends data over multiple subcarriers in parallel, and OFDM is one of the multi-carrier waveforms with orthogonal subcarriers. As illustrated in Fig.~\ref{WaveformDesign}(c), an element-wise-division method can be adopted by the OFDM-sensing receiver to eliminate the influences of transmitted symbols in the frequency domain~\cite{SturmWaveform2011}. Note that the element-wise-division method is also readily applicable to O-ISAC, and therefore an O-ISAC system based on DC-biased optical OFDM (DCO-OFDM) is demonstrated by numerical simulations as follows.

To evaluate communication and sensing performances of the proposed O-ISAC system, $1{\times}10^4$ Monte-Carlo simulations are conducted to obtain the bit error rate (BER) for communication and the root mean square error (RMSE) for target distance estimation, as displayed in Figs.~\ref{Simulation}(a) and~\ref{Simulation}(b), respectively. Each OFDM frame contains $32$ OFDM symbols, while each OFDM symbol consists of $256$ subcarriers with a spacing of $3.9\ \text{MHz}$, yielding a total bandwidth of nearly $1\ \text{GHz}$. For simplicity, all subcarriers adopt the same modulation scheme, i.e., 4QAM or 16QAM. Meanwhile, the sensing receiver utilizes the whole OFDM frame for target distance estimation, yielding a processing gain of 39.1~dB~\cite{SturmWaveform2011}. In addition, the DC bias $b$ is set according to the standard deviation $\sigma$ of the DCO-OFDM signal that is not clipped. Moreover, the channel model in~\cite{LiuPPMMSKSIM2015} is adopted for a 200-m FSO link, which set the attenuation for communication and sensing as -2.2~dB and -23.2~dB, respectively.

Fig.~\ref{Simulation}(a) indicates that the BER is affected by both the DC bias and the noise at the receiver. For an insufficient DC bias, the signal is severely distorted by the non-negative clipping, which deteriorates the communication performance even at the high-SNR region. On the contrary, an exorbitant DC bias reduces the energy efficiency, since the DC bias does not carry any communication data. Likewise, the RMSE can converge to an asymptotic region, as shown in Fig.~\ref{Simulation}(b). The sensing receiver shows superior resistance to the non-linear distortion brought by an insufficient DC bias, as a declined DC bias leaves more power to resist larger channel attenuation for sensing.

\begin{figure*}[tp]
    \begin{center}
        \includegraphics[width=0.96\textwidth]{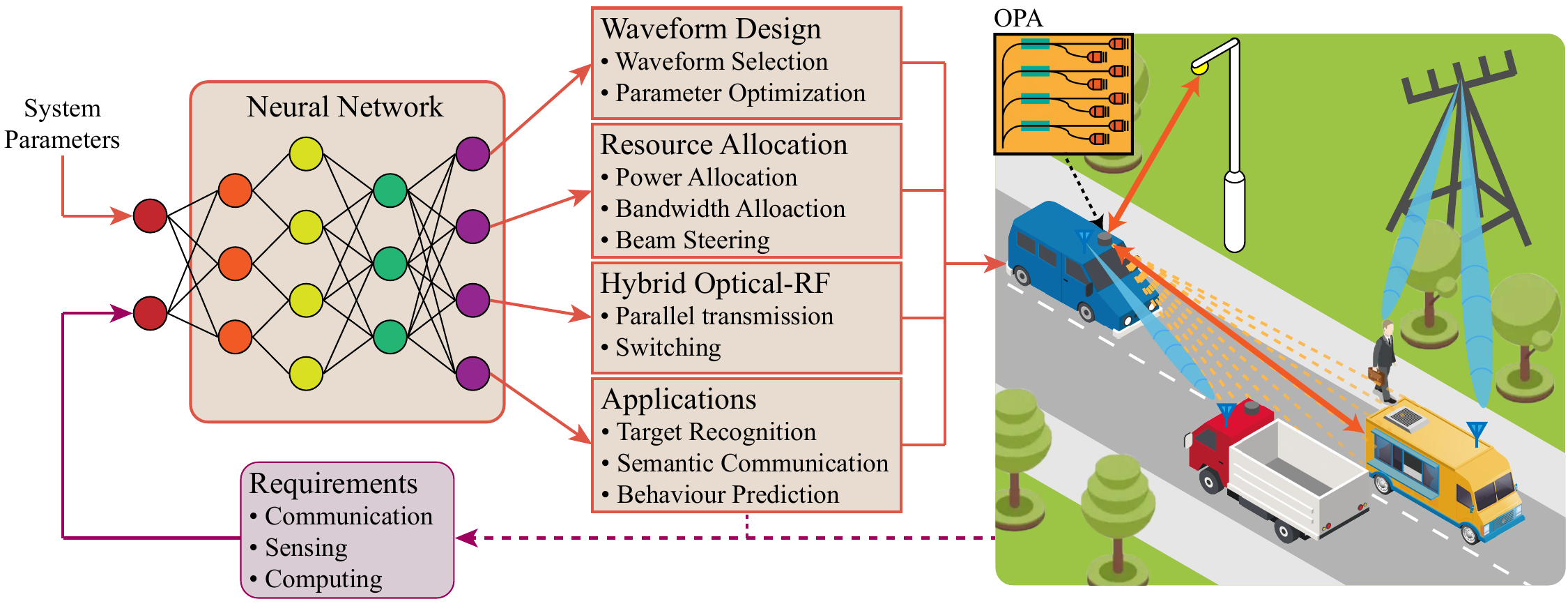}        
        \caption[short]{Future trends of O-ISAC integrated with other emerging technologies.}
        \label{FutureTrends}
    \end{center}
\end{figure*}

\subsection{Resource Allocation and Waveform Optimization}
Once the prototype waveform is selected for O-ISAC, it can be further optimized to achieve superior communication and sensing performances. Taking the LFM-CPM waveform in Subsection~\ref{wav:constant_modulus} as an example, optimization can be conducted on critical parameters like the chirp rate of LFM, the modulation index, and the modulation order of CPM. For a fixed modulation order, the optimal values of chirp rate and modulation index are restricted on an ellipse determined by the sensing constraint, and an elliptical search algorithm is proposed to obtain these optimal values~\cite{WenFSOLFMCPM2023}. Once critical parameters are optimized, the optimal LFM-CPM waveform is also attained.

DCO-OFDM serves as another example for waveform optimization, which achieves the optimal waveform design by power allocation. On one hand, Fig.~\ref{Simulation} indicates that a trade-off exists between the DC bias and the power on other subcarriers under the total power constraint. A quasi-concave relationship between the DC bias and the SNR on other subcarriers is revealed in~\cite{LingOptimizationDCOOFDM2016}, based on which the Newton method can be adopted to obtain the optimal DC bias for flat channels. On the other hand, once the Cramèr-Rao Bound (CRB) is selected as the performance metric, the optimal power allocation for sensing is to allocate all the power to the highest subcarrier. Since the optimal power allocation for communication follows a water-filling form in general, the power allocation on other subcarriers also embodies the trade-off between communication and sensing functionalities. Thereby, to obtain the optimal DCO-OFDM waveform for O-ISAC, a joint optimization problem can be formulated to optimize the DC bias and the power allocation for subcarriers simultaneously.

\section{Future Trends and Challenges for O-ISAC}
As a concept in the ascendant, O-ISAC is confronted with both opportunities and challenges. Advanced hardware and novel performance metrics are in urgent need, while emerging technologies like the hybrid system and deep learning (DL) can also be integrated with O-ISAC. To depict the future trends and challenges, a deep-learning-based O-ISAC system is illustrated in Fig.~\ref{FutureTrends}, while several potential directions and their corresponding challenges are discussed in the folloing subsections.

\subsection{Advanced Hardware for O-ISAC}
The widespread applications of O-ISAC rely on superior durability and miniaturization of optical hardware, which demands the replacement of bulky mechanical components in an optical system. Towards this end, micro-electro-mechanical system, liquid crystal, and optical phased array (OPA) have gained much interest, among which OPA is an attractive scheme for both solid-state laser radars and FSO communication. OPA achieves beam steering by tuning the phases of an optical antenna array, which alters the phase front of the emitted optical beam. An OPA-based laser radar and FSO communication demonstrator is presented by~\cite{PoultonOPA2019}, which achieves velocity extraction and chip-to-chip FSO communication at a distance of nearly 200 m and up to 50 m, respectively. Although advanced hardware contributes to the evolution of O-ISAC systems, theoretical models are urgently required to bridge the gap between novel hardware implementation and optimal system design.

\subsection{Performance Metrics for O-ISAC}
Owing to narrow beams and LoS channels, the received signal for an O-ISAC sensing receiver generally comes from the reflection of a point target, in contrast to the common extended target in RF-ISAC. This difference causes the invalidity of plentiful performance metrics dedicated to multi-target scenarios, like peak-to-sidelobe ratio and integrated-sidelobe ratio. Moreover, establishing an information-theory foundation for sensing is another urgent task for O-ISAC, which contributes to a unified performance metric in joint optimization. The main challenge lies in non-linear distortions of the O-ISAC signal, which yields complicated expressions for performance metrics~\cite{LingOptimizationDCOOFDM2016}. A concise relationship between the waveform design and performance metrics deserves more consideration in future research of O-ISAC.

\subsection{Hybrid RF and Optical ISAC}
Although O-ISAC possesses superior anti-interference abilities, it is also vulnerable to obstacles that cause blind areas, which can be covered by RF-ISAC instead. Furthermore, the delay-doppler analysis enables RF-ISAC to sense multiple moving targets simultaneously, while IM/DD-based O-ISAC cannot extract velocity information directly. Therefore, O-ISAC and RF-ISAC should become powerful complements to each other, yielding a hybrid RF and optical ISAC system illustrated in Fig.~\ref{FutureTrends}. A hybrid RF and optical ISAC system can be implemented in a parallel-transmission-based approach, where power and bandwidth are allocated to both RF-ISAC and O-ISAC simultaneously. Besides, the hybrid system can also be implemented by switching between RF-ISAC and O-ISAC according to channel states and user requirements. The critical problem to be solved is to unify the performance metrics of both RF-ISAC and O-ISAC, through which joint optimization can be conducted to achieve the global optimal design for the hybrid system.

\subsection{Deep Learning for O-ISAC}
As the scenarios and requirements of O-ISAC are increasingly complicated, the end-to-end optimization method of DL can be introduced to enhance the performances of O-ISAC. DL can serve the waveform design and resource allocation process of O-ISAC instead of conventional optimization algorithms. In addition, DL can also provide functionalities of target recognition, behaviour prediction, and semantic communication, which are not included in traditional ISAC. However, the requirements for high-quality datasets and tremendous computing resources are not negligible when designing a DL-based O-ISAC system. One approach to resolving these problems is to conduct training and deduction at computing centers, which adds extra payloads of data transmission. Therefore, while a conventional O-ISAC system mainly focuses on the trade-off between communication and sensing, computing will also become progressively more important in a DL-based O-ISAC system.

\section{Conclusions}
In this article, the generalized system structure of O-ISAC was introduced, based on which three appealing advantages of FSO systems were presented, i.e., increasing communication rate, enhancing sensing precision, and reducing interference. Additionally, waveform design, as the foundation of O-ISAC, was discussed comprehensively based on the commonly adopted pulsed waveform, constant-modulus waveform, and multi-carrier waveform. Furthermore, future trends and challenges for O-ISAC were also put forward, which called for the integration of several cutting-edge technologies with O-ISAC. By addressing these issues, O-ISAC is expected to become a powerful complement to RF-ISAC and also a crucial enabler for numerous applications that will change lives in the near future.

\bibliographystyle{IEEEtran}
\bibliography{Ref.bib, IEEEabrv.bib}

\end{document}